**Title**: Machine Learning Algorithms for Predicting in-Hospital Mortality in Patients with ST-Segment Elevation Myocardial Infarction

Authors: Ding Tao[1, *], MSc; Chen Liu[2], MSc; Shihan Wan[1], MSc

Affiliations:

[1]School of Data Science, The Chinese University of Hong Kong, Shenzhen, P.R. China
[2]Department of Statistics, University of Michigan, Ann Arbor, USA

[*]**Corresponding Author**: School of Data Science, The Chinese University of Hong Kong, Shenzhen, P.R. China

**Abstract**

**Background and purpose:** Acute myocardial infarction (AMI) is one of the most severe manifestation of coronary artery disease. ST-segment elevation myocardial infarction (STEMI) is the most serious type of AMI. We proposed to develop a machine learning algorithm based on the home page of electronic medical record (HPEMR) for predicting in-hospital mortality of patients with STEMI in the early stage.

**Methods**: This observational study applied clinical information collected between 2013 and 2017 from 7 tertiary hospitals in Shenzhen, China. The patients' STEMI data were used to train 4 different machine learning algorithms to predict in-hospital mortality among the patients with STEMI, including Logistic Regression, Support Vector Machine, Gradient Boosting Decision Tree, and Artificial Neuron network.

**Results**: A total of 5865 patients with STEMI were enrolled in our study. The model was developed by considering 3 types of variables, which included demographic data, diagnosis and comorbidities, and hospitalization information basing on HPEMR. The association of selected features using univariant logistic regression was reported. Specially, for the comorbidities, atrial fibrillation (OR: 11.0; 95% CI: 5.64 - 20.2), acute renal failure (OR: 9.75; 95% CI: 3.81 - 25.0), type 2 diabetic nephropathy (OR: 5.45; 95% CI: 1.57 - 19.0), acute heart failure (OR: 6.05; 95% CI: 1.99 - 14.9), and cardiac function grade IV (OR: 28.6; 95% CI: 20.6 - 39.6) were found to be associated with a high odds of death. Within the test dataset, our model showed a good discrimination ability as measured by area under the receiver operating characteristic curve (AUC; 0.879) (95% CI: 0.825 - 0.933).

**Conclusions**: The model based on machine learning from HPMER offer an alternative approach to promptly predict in-hospital mortality among the patients with STEMI. Our findings enable to facilitate quick risk assessment when the patients are admitted owing to STEMI at the first medical contact.

*Introduction*

Acute myocardial infarction (AMI) is one of the most severe manifestation of coronary artery disease. Despite substantial improvements in prognosis over the past decade, AMI remains a leading cause of morbidity and mortality worldwide, accounting for 2.4 million deaths in the United States and more than 4 million deaths in Europe and North Asia, and about one third of deaths come from developed countries every year (Reed et al., 2017; Yeh et al., 2010). According to the existence of ST-segment elevation on electrocardiogram (ECG), AMI can be divided into ST-segment elevation myocardial infarction (STEMI) and non-ST segment elevation myocardial infarction (NSTEMI) (Vogel et al., 2019). At least in the short term, STEMI was proved to be more hazardous, and linked with a higher mortality compared to NSTEMI(Aude et al., 2013). Complete thrombotic occlusion of epicardial coronary atherosclerotic plaque is the leading cause of STEMI. Several previous studies had highlighted a fall in acute and long-term mortality following in parallel with the improved management of patients with STEMI. The most revolutionary change was the development of primary percutaneous coronary intervention (PCI) as the favoured reperfusion strategy for patients with STEMI. Additional refinements in PCI techniques and STEMI management further improved clinical outcomes, resulting in markedly low mortality rates in the past decades. The past years witnessed great strides regarding to STEMI management and major advances in the treatment of patients with AMI. However,the incidence of AMI and its associated mortality remains substantial, with an immense impact on patients and healthcare systems.

The mortality in STEMI patients is affected by a coalescence of several factors, including the physical condition of patients, the access to interventional treatment and the absence of an organized protocol for STEMI management. Numerous evidence suggested that poor access to primary PCI and prolonged interval time (<120 min) from first medical contact to PCI were responsible, in part, for significantly worse clinical outcomes in patients with STEMI. In addition to this, the treatment strategy, largely depending on the time delay to PCI, plays a nonnegligible role in the clinical outcomes of patients with STEMI.

A randomized trial demonstrated that early coronary angiography could effectively reduce the occurrence of adverse events in high-risk patients with AMI (Deharo et al., 2017). In this regard, understanding factors associated with in-hospital mortality after STEMI can guide management decisions at an early stage as well as provide valuable prognostic information not only for clinicians but also for patients and their families. In addition, models based on risk help account for case-mix and other relevant factors when attempting to compare hospital outcome performance.

Therefore, a prediction of the severity and prognosis is vital for identifying patients at high risk and providing intensive treatment and monitoring. Currently, several risk scores have been developed, based on readily identifiable parameters in the acute phase before reperfusion. For instance, the Global Registry of Acute Coronary Events (GRACE) risk score is recommended for risk assessment and adjustment (Granger et al., 2003), Canadian ACS risk score (Huynh et al., 2013), and Acute Coronary Treatment and Intervention Outcomes Network (ACTION) (McNamara et al., 2016) had been successively introduced to estimate the mortality risk in patients with AMI.

Although these models were validated and are commonly accepted tools, concerns have been raised recently because most traditional risk stratifications were developed many years ago using randomized controlled trial (RCT) data before the introduction of drug-eluting stents and newer generation antiplatelets. In recent years, several models had been introduced to predict AMI whereby using data-driven methods (Karabağ et al., 2018; Song et al., 2019; Wang et al., 2021; Wu et al., 2021). Among these prediction model, machine learning (ML) has been widely acknowledged for its high performance in outcomes of prediction. Most previous published studies were intensively concentrated the in-hospital mortality of NSTEMI and conducted in Europe, United States or Canada. In addition, these studies aimed to predict the mortality risk mainly through demographic data, lifestyle, physical examinations like ST-segment depression on ECG, and laboratory test. Furthermore, the high computation power and many clinical predictors, which are difficult to extract from the electronic medical records, limit the use of prediction models using deep learning algorithms in clinical practice. However, HPMER usually includes the patients' basic information, such as socioeconomic status, the patients' age, occupation, medical payment style, diagnostic information, and comorbidities and can be obtained at the first medical contact of the patients before obtaining laboratory testing.

To the best of our knowledge, previous published studies on its prediction for patients with STEMI are still limited, especially for patients outside the non-European countries. Thus, we proposed to set up a model using ML for predicting the in-hospital mortality in patients with STEMI through analyzing the HPMER data from 7 hospitals in Shenzhen, China.

## *Methods*

**Data Collection**

The data used in our study was collected from HPEMR in the highly standardized EMR system of 7 hospitals in Shenzhen, China. Specifically, structured data were collected for all patients (n = 18,181) hospitalized for acute myocardial infarction (the International Classification of Disease, Tenth Revision (ICD-10) codes: I21) between 2014 and 2017. For model development, HREMR data, including patients' demographic information, medical history, diagnosis, comorbidities, and hospitalization information, were collected and analyzed. The exclusion criteria (Figure 1) were as following: (1) main diagnosis was not STEMI according to the ICD-10 codes; (2) ages were below 18 above 90; (3) the in-hospital or discharge records with missing items; (4) the patients were discharged or transferred without the clinicians' advice. After applying the exclusion criteria, 5,865 patients were eligible for our study. Ethics approval was provided by all the 7 hospitals in Shenzhen, China. Conduction of this audit program had been performed in accordance with the Declaration of Helsinki, and reported to the 7 involved hospitals Ethics Commission. In addition, approval was obtained beforehand. All of the data were anonymized before processing, hence, participant's privacy and confidentiality were protected.

**Outcome Variable**

The outcome variable of the our study was whether the patients died during hospitalization days. In HPEMR, the ways of departure from hospital included being discharged or transferred, and died. According to a retrospective study on the relationship between discharge type and condition (Pages et al., 1998), the patients discharged against advice had a shorter length of staying, 30 days higher of hospitalization rates, and severe symptoms at discouragement. Also, the patients transferred to other hospitals typically had severe symptoms. These patients were excluded from our cohort study. In the end, our study contains 5,655 patients discharged or transferred with medical advice, and 210 patients died in the hospitals.

**Model Framework**

*Framework Procedure*

The framework of our machine learning algorithm is presented in Figure 2. It shows the whole process of the experiment, including data extraction, data preprocessing, modeling, and model evaluation. We obtained three categories of data from the HPEMR, namely, demographic, diagnosis and comorbidity, and hospitalization information. Then, the eligible cohort was selected according to the exclusion criteria. The next step is to do the preprocessing. We filtered features according to univariate analysis and built some new valuable features relaying on existed data. The data was splitting into two parts. Data from 2014 to 2016 was used as training and validation sets (n = 3860), and data from 2017 was treated as testing sets (n = 2005). Four algorithms were trained in the modeling stage: Logistic Regression (LR), Support Vector Machine (SVM), Gradient Boosting Method (GBDT), and Artificial Neuron Network (ANN). Algorithm searches for the best hyperparameters using grid search to get a high model performance in validation set. We used 10-fold cross-validation to train all the learning algorithms. Once we have tuned the optimal hyperparameters in the training phase, the model will be tested on the datasets from 2017 and output the performance results.

*Data Preprocessing*

In data preprocessing step, we mainly focused on screening the patients' comorbidities data. We regarded comorbidities as precious indicators in the model for early prediction. According to the frequency of commodities, the less common comorbidities in all patients were eliminated, and some general comorbidities were screened out because they may not be differentiated from the patients. Total of 16 comorbidities were selected at this step. As the study by Liu (Lin et al., 2020) stated that the admission time being at a holiday had a certain impact on the outcome of patients with AMI in China, so we took this factor (whether the admission time is a holiday or not) into account in our mode. Briefly, most of the variables were considered, then filtered out, not related, and confounding variables. Finally, the features used to build the algorithm can be grouped to: (1) demographic data, including age, sex, occupation, types of admission, and types of medical payment method; (2) diagnosis and comorbidity information; (3) hospitalization information such as the number of hospitalizations, whether a holiday was at admission and whether the patients' diagnosis and treatment were based on clinical pathway.

*Machine Learning Algorithms*

We compared four machine learning algorithms, which can be grouped into linear models and nonlinear models. Linear model: Logistic Regression (LR) is often used in clinical research because it is easier to explain. It transforms linear regression into a binary classifier with sigmoid function. Support Vector Machine (SVM) (Noble, 2006) divides two categories of samples through hyperplane. Nonlinear model: Gradient Boosting Decision Tree (GBDT) is ensemble of decision trees, which classifies data by establishing decision rules; Artificial neural network (ANN) (Mishra & Srivastava, 2014) extracts essential information from the input with neuron layer, and then uses the features for classification.

*Model Evaluation*

In this study, performance was measured by 4 scores contains Area under the receiver operating characteristic curve (AUC), specificity, sensitivity, and F1-Score. AUC is an aggregate measure of the algorithm's ability to discriminate outcome classes across all possible classification thresholds (Huang & Ling, 2005). At the same time, AUC score is robust to the unbalanced data. Specificity, sensitivity, and F1-Score are also reported to support the performance. Machine learning algorithms were trained and evaluated using Scikit-learn and Matplotlib in Python (3.8).

## Results

*Study Population*

A total of 5,865 patients with STEMI from 2014 to 2017 were included in the database fitting the inclusion criteria. Table 1 shows the descriptive statistics of the cohort. We found there was significant difference in sex, age, type of occupations, type of admission, number of hospitalizations and whether the patients' diagnosis and treatment were based on the clinical pathway between died and survival patients. The average age of patients was 57.5 years (SD = 13.1) in the survival group and 69.7 (SD = 14.0) in the died group. The number of hospitalizations of patients in the died group was higher than that in the survival group (2.19 vs 1.38 days, $p < 0.001$). Odds ratios can reflect the degree of association between exposure factors and risk of death. It can be found that females were associated with a higher odds ratio of death suffering STEMI when compared with the males (OR: 2.62; 95% CI:1.95 - 3.49). As for the occupation, enterprise manager (OR: 4.33; 95% CI: 1.43 - 11.9), jobless (OR: 2.65; 95% CI: 1.4 - 5.49), retired (OR: 4.35; 95% CI: 2.42 - 8.69), and self-employed (OR: 4.32; 95% CI: 1.53 - 11.4) were also associated with a high odds ratio of death. Besides, compared to outpatient admission, the type of emergency (OR: 2.01; 95% CI: 1.33 - 3.18) and transfer from other hospitals (OR: 4.05; 95% CI: 2.07 - 7.73) were associated with a higher odds ratio of death. The patients whose diagnosis and treatment were based on the clinical pathway were associated with lower odds of death (OR: 0.34; 95% CI: 0.26 - 0.45) than those who did not obey the clinical pathway.

*Comorbidities Analysis*

Table 2 represents the descriptive statistics of selected comorbidities. More than half of the chosen comorbidities were significantly associated with the outcome of the patients. The comorbidities include Ventricular aneurysm, Atrial fibrillation, acute renal failure, Cardiac function class III, Type 2 diabetic nephropathy, Acute heart failure, heart failure, and Cardiac function grade IV were associated with high odds of death. Typically, Atrial fibrillation (OR: 11.0; 95% CI: 5.64 - 20.2), acute renal failure (OR: 9.75; 95% CI: 3.81 - 25.0), Type 2 diabetic nephropathy (OR: 5.45; 95% CI:1.57 - 19.0), Acute heart failure (OR: 6.05; 95% CI: 1.99 - 14.9), and Cardiac function grade IV (OR: 28.6; 95% CI: 20.6 - 39.6) were found to be very dangerous comorbidities with high odds of death.

*Performance of the Models*

The efficiency of four different ML models is shown in Table 3, and the model of receiver operating characteristic (ROC) curves is shown in Figure 3. Within testing cohort (n = 2005), 66 died during hospitalization. Logistic Regression was found to be the best performing algorithm to predict in-hospital mortality of patients with STEMI using HPMER. The AUC of logistic regression was 0.879 (95% CI: 0.825 - 0.933). When we set the cut-off value equals to 0.02, Logistic Regression achieved performance with a sensitivity of 81%, a specificity of 88%, and F1-score of 65%. The AUC of GBDT (0.861; 95% CI: 0.803 - 0.917) and ANN (0.865; 95% CI: 0.809 - 0.912) are close to Logistic regression. Generally, LR model achieved the best prediction performance among the machine learning models.4

## Discussion

*Principal Results*

This study established a machine learning prediction model and could be implemented in the early stage to assess the in-hospital mortality risk of patients with STEMI. Our model considered 3 types of variables (demographic data, diagnosis and comorbidities, and hospitalization information basing on HPEMR) that can be obtained in the first medical contact without the need to obtain the laboratory test results and treatment information. Importantly, the model has a good discrimination ability as measured by AUC score (0.879; 95% CI: 0.825 - 0.933). This supports early stratification of high-risk patients and helps to the selection of treatment option. In addition, our model identifies that comorbidities and demographic predictors were highly associated with in-hospital death by univariant analysis.

STEMI, as a life threatening complication of coronary artery disease, has become one of the leading cause of death worldwide. So, it is of great importance to find effective ways to predict the short- and long-term mortality among the patients with STEMI. Currently, ML was suggested to have priority in improving the performance of the prediction model, for it could overcome the limitations of a regression-based risk score system (Lee et al. 2021). It has been used in the risk or mortality prediction in patients with AMI (Austin & Lee, 2011; Barrett et al., 2019; Wu et al., 2021). Recently, some studies focused on the development of the prediction model among the patients with NSTEMI; however, the short-term mortality of NSTEMI is lower than that of STEMI (Ahrens et al., 2019; Deharo et al., 2017; Wang et al., 2021). Therefore, it seems that it is more necessary to develop the mortality prediction model among the patients with STEMI in terms of short-term mortality. In early 20 years ago, Morrow et al. (Morrow et al., 2001) applied TIMI Risk Score to built the prediction model in patients with STEMI, when didn't have drug-eluting stents treatment. Recent studies with regards to predicting in-hospital and long-term mortality in patients with STEMI were reported from other countries, like South Korea and Turkey, rather than China (Karabağ et al., 2018; Lee et al., 2021). So, it lacks sufficient data of mortality prediction model in Chinese patients with STEMI. To bridge the knowledge gap, we found a simple way basing on HPEMR and applied ML to develop the model for the early prediction of in-hospital mortality in patients with STEMI. It can be seen that our prediction model developed using the ML algorithm has a good discrimination ability with an AUC score of 0.879.

In the process of univariant analysis of patient demographic information and comorbidities information, we found several high-risk factors associated with high odds ratio of death. Regarding the occupations, enterprise managers, retirees and freelancers are more dangerous factors than office workers, which may be attributed to the great pressure and uncertainty in their lives. As for type of admissions, it is reasonable that emergency admission and transfer from other hospitals are more dangerous factors than outpatient admission, because they usually have more serious conditions. We obtained similar results with a previous study (Lin et al., 2020) on the factor of whether a holiday at admission is a risk factor. Besides, we found whether the patients' diagnosis and treatment were based on clinical pathway is a significant protective factor, which supports the medical staff to manage patients according to the clinical pathway guidelines of the disease. Regarding the comorbidities, atrial fibrillation, acute renal failure, acute heart failure and cardiac function grade IV were observed to be dangerous factors for patients with STEMI, which can be explained by the fact that these factors themselves have a high mortality rate in cardiovascular disease (Arrigo et al., 2020; Liaño & Pascual, 1996; Ruddox et al., 2017).

## Limitations

This study had several limitations. First, our study cohort came from one city in China and other cities are needed in the future studies to improve the generalization performance. Second, we only considered in-hospital mortality. Although STEMI has a high mortality rate in the short-term, it also has a great chance of recurrence and death after hospitalization (Aude et al., 2013). Long-term data is beneficial to future study to has a long-term risk prediction. Third, HPEMR does not include laboratory test results and treatment information during hospitalization. Collecting those variables may help to improve model performance in the future research.

## Conclusions

In summary, our study provides a simple data-driven approach using HPEMR data to predict in-hospital mortality of STEMI patients in China. On the other hand, the prediction can complete when the first medical contact.

## Abbreviations:
AMI: Acute myocardial infarction; STEMI: ST-segment elevation myocardial infarction;
HPEMR: home page of electronic medical record; NSTEMI: non-ST segment elevation myocardial infarction; ICU: Intensive care unit; ECG: electrocardiogram; PCI: percutaneous coronary intervention; RCT: randomized controlled trial; ML: machine learning; ROC: Receiver operating curve; AUC: Area under the curve; SVM: support vector machine; LR: logistic regression; GBDT: gradient boosting decision tree; ANN: artificial neuron network; PPV: Positive predictive value; TPR: True positive rate.

## Declarations

**Ethics approval and consent to participate:**
The Chinese University of Hong Kong 's institutional review board approved the study protocol and waived the requirement for participants' informed consent owing to the infeasibility of acquiring consent for medical record data.

**Acknowledgements:**
Not applicable

**Authors' contributions:**
Ding contributed to the design of the study, analysis the data and write manuscript. Chen contributed to data modeling and analysis. Shihan and Chen contributed to the writing of the manuscript. All authors read and approved the final manuscript.

**Funding:**
Study was funded by The Chinese University of Hong Kong, Shenzhen.

**Consent for publication:**
Not applicable.

**Availability of data and materials:**
The datasets generated and/or analyzed during the current study are not
publicly available but are available from the corresponding author on reasonable request.

**Competing interests:**
We declare that we have no financial and personal relationships with other people or organizations that can inappropriately influence our work, there is no professional or other personal interest of any nature or kind in any product, service and/or company that could be construed as influencing the position presented in, or the review of, the manuscript entitled, "Machine Learning Algorithms for Predicting in-Hospital Mortality in Patients with ST-Segment Elevation Myocardial Infarction".

**Disclosure statement:**
The authors report there are no competing interests to declare.

**Figure legends**

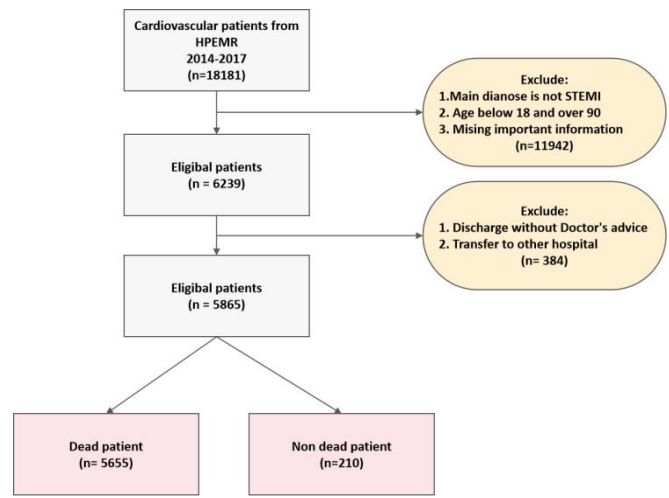

Fig. 1 The inclusion and exclusion criteria of this study.

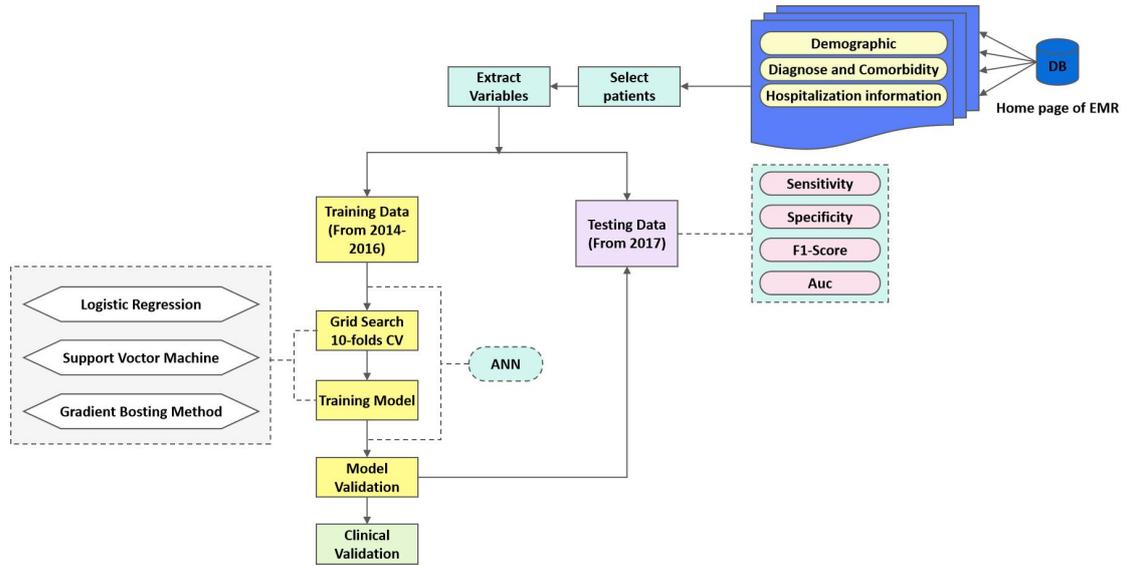

Fig. 2 Framework of the machine learning algorithm. This diagram depicts the process of the study.

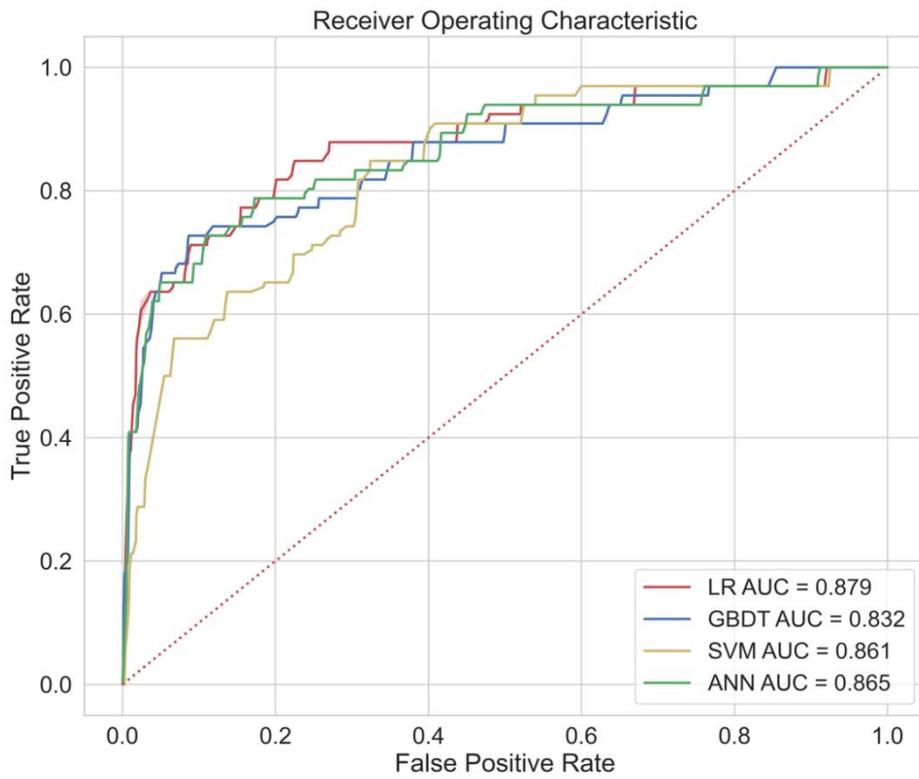

Fig. 3 ROC curves of the in-hospital mortality prediction models.

Table 1 Baseline characteristics of patients who died vs survived

| | Non-Death (N = 5655) | Death (N = 210) | OR | P. ratio | P-value (overall) |
|---|---|---|---|---|---|
| Sex (%): | | | | | <0.001 |
|     Male | 4681 (82.8%) | 136 (64.8%) | Ref. | Ref. | |
|     Female | 974 (17.2%) | 74 (35.2%) | 2.62 [1.95;3.49] | <0.001 | |
| Age | 57.5 (13.1) | 69.7 (14.0) | 1.08 [1.06;1.09] | <0.001 | <0.001 |
| Occupation (%): | | | | | . |
|     Office Worker | 622 (11.0%) | 11 (5.24%) | Ref. | Ref. | |
|     Factory-hand | 383 (6.77%) | 5 (2.38%) | 0.75 [0.23;2.11] | 0.598 | |
|     Farmer | 289 (5.11%) | 4 (1.90%) | 0.80 [0.21;2.40] | 0.708 | |
|     Enterprise manager | 79 (1.40%) | 6 (2.86%) | 4.33 [1.43;11.9] | 0.012 | |
|     retirees | 1256 (22.2%) | 98 (46.7%) | 4.35 [2.42;8.69] | <0.001 | |
|     Jobless | 864 (15.3%) | 41 (19.5%) | 2.65 [1.40;5.49] | 0.002 | |
|     Professional Technicians | 476 (8.42%) | 7 (3.33%) | 0.84 [0.30;2.17] | 0.720 | |
|     Self-employed | 92 (1.63%) | 7 (3.33%) | 4.32 [1.53;11.4] | 0.007 | |
|     Other | 1594 (28.2%) | 31 (14.8%) | 1.09 [0.56;2.29] | 0.809 | |
| Type of Admission (%): | | | | | <0.001 |
|     Outpatient | 1213 (21.5%) | 24 (11.4%) | Ref. | Ref. | |
|     Emergency | 4242 (75.0%) | 170 (81.0%) | 2.01 [1.33;3.18] | 0.001 | |
|     Transfer | 200 (3.54%) | 16 (7.62%) | 4.05 [2.07;7.73] | <0.001 | |
| Number of Hospitalizations | 1.38 (1.53) | 2.19 (3.54) | 1.13 [1.07;1.18] | <0.001 | 0.001 |
| Medical Payment Method (%): | | | | | 0.638 |
|     Medical insurance for urban residents | 396 (7.00%) | 10 (4.76%) | Ref. | Ref. | |
|     Medical insurance for urban workers | 1970 (34.8%) | 71 (33.8%) | 1.41 [0.75;2.94] | 0.300 | |
| Cash | 2591 (45.8%) | 101 (48.1%) | 1.52 [0.83;3.15] | 0.188 | |
| At one's own expense | 373 (6.60%) | 17 (8.10%) | 1.79 [0.82;4.14] | 0.146 | |
| Other | 325 (5.75%) | 11 (5.24%) | 1.34 [0.55;3.28] | 0.516 | |
| Is a holiday at Admission (%): | | | | | 0.149 |
| No | 4197 (74.2%) | 146 (69.5%) | Ref. | Ref. | |
| Yes | 1458 (25.8%) | 64 (30.5%) | 1.26 [0.93;1.70] | 0.032 | |
| Whether through clinical pathway: | | | | | <0.001 |
| No | 1453 (25.7%) | 106 (50.5%) | Ref. | Ref. | |
| Yes | 4202 (74.3%) | 104 (49.5%) | 0.34 [0.26;0.45] | <0.001 | |

Table 2 The association of selected comorbidities using univariate logistic regression in training data.

|  | Non-Death (N = 5655) | Death (N = 210) | OR | P-value |
|---|---|---|---|---|
| Ventricular Aneurysm | 53 (0.9%) | 5 (2.36%) | 2.74 [0.93;6.31] | 0.038 |
| Atrial Fibrillation | 38 (0.64%) | 14 (6.60%) | 11.0 [5.64;20.2] | <0.001 |
| Acute Renal Failure | 24 (0.41%) | 5 (2.36%) | 9.75 [3.81;25.0] | <0.001 |
| Cardiac Function Grade 3 | 144 (2.44%) | 10 (4.72%) | 2.00 [1.04;3.86] | 0.039 |
| Hypertension Grade 1 | 271 (4.60%) | 5 (2.36%) | 0.52 [0.18;1.14] | 0.123 |
| Hypertension Grade 2 | 547 (9.28%) | 11 (5.19%) | 0.53 [0.29;0.99] | 0.033 |
| Cardiac Function Grade 1 | 4329 (73.4%) | 62 (29.2%) | 0.11 [0.08;0.14] | <0.001 |
| Cardiac Function Grade 2 | 334 (5.66%) | 6 (2.83%) | 0.50 [0.19;1.03] | 0.063 |
| Acute Heart Failure | 24 (0.41%) | 5 (2.36%) | 6.05 [1.99;14.9] | 0.003 |
| Heart Failure | 35 (0.59%) | 4 (1.89%) | 3.33 [0.97;8.47] | 0.025 |
| Type 2 Diabetes | 407 (6.90%) | 6 (2.83%) | 0.40 [0.16;0.84] | 0.012 |
| Type 2 Diabetic Nephropathy | 17 (0.29%) | 3 (1.42%) | 5.45 [1.57;19.0] | 0.008 |
| Cerebral Infarction | 42 (0.71%) | 2 (0.94%) | 1.42 [0.21;4.67] | 0.650 |
| Cardiac Function Grade IV | 133 (2.26%) | 84 (39.6%) | 28.6 [20.6;39.6] | <0.001 |
| Coronary Artery Stenosis | 93 (1.58%) | 6 (2.83%) | 1.86 [0.71;3.96] | 0.186 |

Table 3 The model performance

| Classifier | AUC | Sensitivity | Specificity | F1-Score |
|---|---|---|---|---|
| Logistic Regression | 0.879 (0.825, 0.933) | 0.81 | 0.88 | 0.65 |
| Support Vector Machine | 0.832 (0.772, 0.894) | 0.82 | 0.87 | 0.49 |
| Gradient Boosting Tree | 0.861 (0.803, 0.917) | 0.78 | 0.85 | 0.57 |
| Artificial Neuron Network | 0.865 (0.809, 0.912) | 0.79 | 0.84 | 0.62 |